# Frequency comb spectroscopy

Nathalie Picqué [1,*] and Theodor W. Hänsch [1,2]

1. Max Planck Institute of Quantum Optics, Hans-Kopfermannstr. 1, 85748 Garching, Germany
2. Ludwig-Maximilian University, Faculty of Physics, Schellingstr. 4, 80799 Munich, Germany
* Corresponding author: nathalie.picque@mpq.mpg.de

**Abstract**
A laser frequency combs is a broad spectrum composed of equidistant narrow lines. Initially invented for frequency metrology, such combs enable new approaches to spectroscopy over broad spectral bandwidths, of particular relevance to molecules. With optical frequency combs, the performance of existing spectrometers, such as Michelson-based Fourier transform interferometers or crossed dispersers, involving e.g. virtual imaging phase array (VIPA) étalons, is dramatically enhanced. Novel types of instruments, such as dual-comb spectrometers, lead to a new class of devices without moving parts for accurate measurements over broad spectral ranges. The direct self-calibration of the frequency scale of the spectra within the accuracy of an atomic clock and the negligible contribution of the instrumental line-shape will enable determinations of all spectral parameters with high accuracy for stringent comparisons with theories in atomic and molecular physics. Chip-scale frequency-comb spectrometers promise integrated devices for real-time sensing in analytical chemistry and biomedicine. This review article gives a summary of advances in the emerging and rapidly advancing field of atomic and molecular broadband spectroscopy with frequency combs.

## 1. Introduction

A frequency comb is a spectrum of phase-coherent evenly spaced narrow laser lines (Fig. 1a). Frequency combs[1] have revolutionized time and frequency metrology in the late 1990's by providing rulers in frequency space that measure large optical frequency differences and/or straightforwardly link microwave and optical frequencies. Very rapidly, frequency combs have found applications beyond the original purpose. For instance, they provide long-term calibration of large astronomical spectrographs[2]; by enabling the control of the relative phase between the envelope and the carrier of ultrashort pulses, they have become a key to attosecond science[3]; low-noise frequency combs of high repetition frequency benefit radio-frequency arbitrary waveform generation and optical communications[4]. In the present review article, we focus on their impact in spectroscopy where the frequency comb is used to directly excite or interrogate the sample. This field is sometimes called direct frequency comb spectroscopy, or broadband spectroscopy with frequency combs. In the following, we coin it frequency comb spectroscopy. We do not discuss the applications to precision spectroscopy where the comb is used as a frequency ruler, also called comb-assisted or comb-referenced spectroscopy, for which the reader can refer e.g. to Refs. [5,6]. In addition, we explicitly concentrate on a selection of comb synthesizers and techniques where the comb structure is exploited, even though the field of spectroscopy with broadband coherent sources of other types is also exhibiting significant progress.

This review article presents a summary of advances in the growing field of frequency comb spectroscopy and its emerging applications. The advent of frequency comb spectroscopy brings a set of new tools to spectroscopy in all phases of matter. In the simplest approach (Fig. 1b), a frequency comb excites and interrogates the sample. The spectral response of the sample, due e.g. to linear absorption or to a nonlinear phenomenon, may span the entire bandwidth of the comb and therefore a spectrometer is required (apart from exceptions, presented e.g. in sections 3.1 and 3.2). Despite daunting technical challenges, the last decade has witnessed remarkable progress in laser frequency-comb sources dedicated to broadband spectroscopy, especially in the molecular-fingerprint mid-infrared (2-20 μm) region and the ultraviolet range (<400 nm). Existing spectrometers and spectrometric techniques are adapted and improved to resolve individual comb lines, while comb-enabled instruments are explored. First studies suggest new opportunities for exploring atomic and molecular structure and dynamics. Meanwhile, spectroscopic sensors, with enhanced capabilities to probe a variety of environments, are devised.

Frequency comb spectroscopy has been the first application of a frequency comb in the 1970's. The regular train of picosecond pulses of a mode-locked laser was then shown to be usable for Doppler-free two-photon excitation spectroscopy of simple atoms [7,8], with the same efficiency as a continuous-wave laser of the same average power [8,9]. Although the narrow spectral span at the output of a picosecond laser would not allow absolute referencing, the role of the carrier-envelope offset frequency in the comb spectrum was already understood at that time [10]. In the 2000's, after self-referencing[11] of femtosecond laser synthesizers established the power of frequency-comb techniques for comb-assisted precision spectroscopy, frequency comb spectroscopy began to attract the interest of a few research groups [12-20]. Stimulated by a number of intriguing proof-of-principle demonstrations, the field has grown to a hot research topic and, with new research groups constantly getting involved, novel clever and unforeseen applications emerge. Most of the time, frequency comb spectroscopy implies spectroscopy over a broad spectral bandwidth, of particular relevance to molecular spectroscopy.

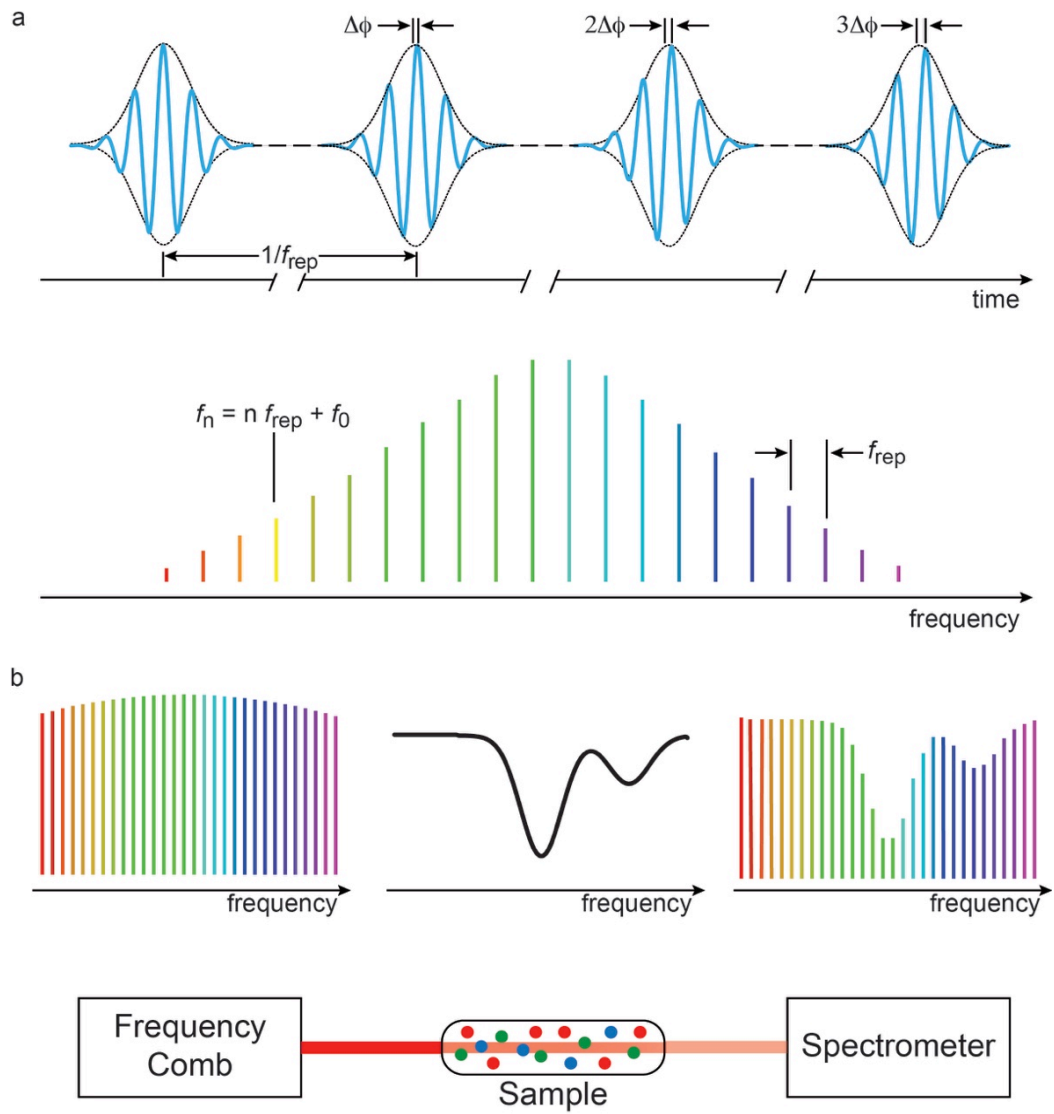

**Figure 1. Principle of a frequency comb and sketch of a simple experiment of frequency comb spectroscopy.**
**a.** Time domain representation of the train of ultra-short pulses of period $1/f_{rep}$ at the output of a mode-locked laser and the corresponding spectrum of narrow lines of a frequency comb. The phase-shift $\Delta\phi$ of the carrier of the wave relatively to the envelope of the pulses induces a translation $f_0 = f_{rep} \Delta\phi/2\pi$ of all the lines in the spectrum from their harmonic frequencies $nf_{rep}$.
**b.** In the simplest experiment of frequency comb spectroscopy, a frequency comb as a broadband light source interrogates an absorbing sample and a spectrometer analyzes the transmission spectrum.

## 2. Frequency-comb light sources for spectroscopy.

<u>2.1 General characteristics.</u>
Often, a comb is generated by a mode-locked laser system. In the time domain (Fig. 1a), a train of ultrashort pulses is emitted at the output of the cavity. The period of the envelope of the pulses, $1/f_{rep}= L/v_g$ , corresponds to the round-trip time inside a laser cavity of round-trip length $L$ and group velocity of the light $v_g$. Due to dispersion inside the cavity, a pulse-to-pulse phase-shift $\Delta\phi$ between the carrier and the envelope of the electric field of the pulses is observed. In the frequency domain, the associated spectrum is composed of a discrete set of evenly spaced narrow lines with frequencies $f_n$ that can be written $f_n$= n $f_{rep}$ + $f_0$, where n is a large integer, $f_{rep}$ is the repetition frequency of the envelope of the pulses and $f_0$ is the carrier-envelope offset frequency, related to the phase-shift $\Delta\phi$ by the relationship $f_0 = f_{rep} \Delta\phi/2\pi$  (Fig. 1a).

In the following, we discuss some common features that apply to many experiments in frequency comb spectroscopy.

All regions of the electro-magnetic spectrum offer interesting opportunities for spectroscopy. However, many applications involve interrogating strong rotational or ro-vibrational transitions in molecules or electronic (rovibronic) transitions in atoms (molecules). Therefore, the targeted regions are the sub-millimetre (also called THz. Wavelength range: 100μm-1mm; frequency range: 0.3-3THz), the far- (100-20 μm, 3-15THz), the mid- infrared (2-20 μm, 15-150THz) and the ultra-violet (<400 nm, >750 THz) ranges.

A broad spectral span is advantageous, especially for spectroscopy of molecules. Most spectrometers have a limited dynamic range and therefore combs with a flat intensity spectral distribution are also beneficial. Furthermore, not all applications require self-referencing. Many combs for frequency comb spectroscopy use the output of the laser system, possibly moderately broadened in a nonlinear fibre, rather than the octave-spanning spectrum, used for instance in *f*-2*f* interferometers, that suffers from strong variations of intensity across the span.

Experiments of linear spectroscopy usually do not require a large average power, as the single photodiode or the camera of the spectrometer can only admit a limited power. Therefore, many experiments of absorption spectroscopy harness an average power on the μW- or mW-scale. Conversely, experiments involving the generation of nonlinear phenomena at the sample often require high pulse energies and therefore higher average power, up to the W level, are reported.

The most suited repetition frequency strongly depends on the type of spectroscopy and the type of sample. Ideally, the repetition frequency should be similar to the desired spectral resolution. Furthermore, resolving the individual comb lines brings the possibility of defining the spectral elements more precisely than by their spacing and of directly calibrating the frequency scale, rather than relying on spectral reference lines. For samples in the condensed phase, a line spacing in the range 50-500 GHz may be desirable. Conversely, for the study of Doppler-broadened transitions of light gas molecules at room temperature in the mid-infrared region, a repetition frequency in the range 50-200 MHz is advantageous. Doppler-free spectroscopy, heavy molecules, or cold samples may require an even narrower spacing. However, few spectrometers have a sufficient resolution to resolve such comb lines directly.

Figure 2 provides an illustration of the spans and repetition frequencies of selected comb generators across the sub-millimetre and infrared regions.

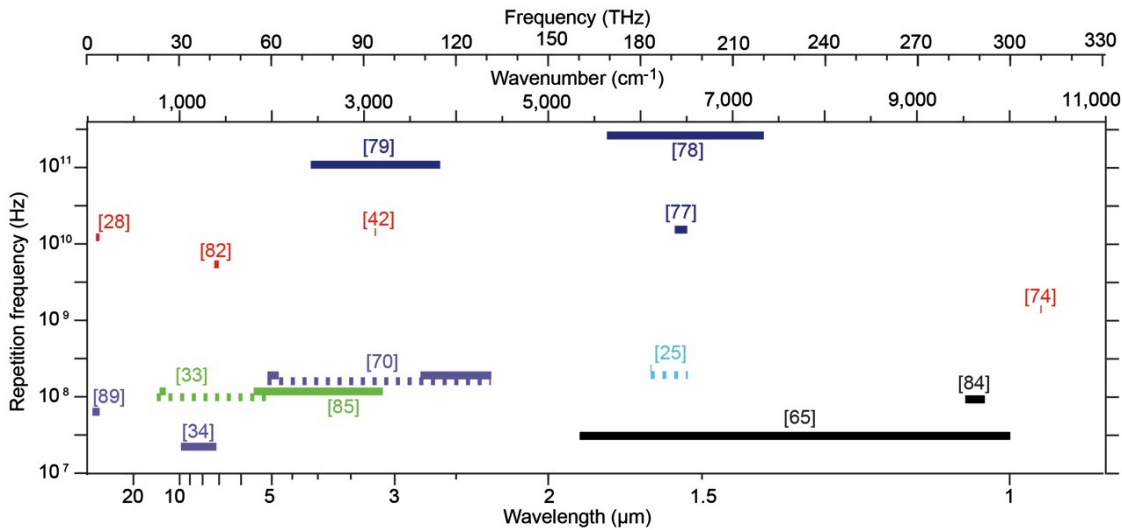

**Figure 2. Spectral coverage with a selection of sources available for frequency comb spectroscopy**

Overview of the spectral spans and repetition frequencies accessible with femtosecond lasers (bar in black), semiconductor lasers (red), microresonator based Kerr combs (dark blue), difference frequency generation (violet), optical parametric oscillators (green) and electro-optic modulators (cyan) from the THz to the near-infrared region. The width of each bar represents the spectral span over which the respective comb emits (dashed line: tuning range, solid line: span for one setting) and the ordinate gives the repetition frequency of the source. The selection, which focuses on comb sources designed for spectroscopy, is intended as an illustration and it does not provide an exhaustive summary of available frequency comb generators.

2.2 Near-infrared region.

Frequency-comb technology in the near-infrared region (800nm–2μm, 150-375 THz) is now mature. Frequency-comb mode-locked lasers are commercially available, the most widespread gain media being Ti:Sa and fibres doped with ytterbium, erbium, thulium or holmium. Recent developments of sources without mode-locked lasers expand the capabilities in terms of repetition frequency, span, compactness etc. Many approaches that have a potential for operation in the mid-infrared region are first tested in the telecommunication region around 1.5 μm (200 THz). Soliton Kerr combs generated in high-quality-factor micro-resonators, as reviewed in ref.[21], provide convenient access to a range of repetition frequencies, from 10 GHz to 1 THz, which is difficult, or even impossible, to reach with conventional mode-locked lasers. The recent demonstration of a battery-operated Kerr frequency-comb generator [22], as well as that of a III-V-on-Si mode-locked laser[23], holds much promise for fully integrated chip-based synthesizers. Frequency-agile combs[24, 25], based on architectures involving one or several electro-optic modulators, exhibit flat-top spectra with a moderate number of lines (up to 10,000) but a freely selectable line spacing and a centre frequency that can be rapidly tuned.

2.3 Sub-millimetre (THz) and far-infrared region.
Frequency-comb sources directly emitting in the THz (1 mm-100 µm, 0.3-3 THz) and far-infrared (100-20 µm, 3-15 THz) regions have long remained inexistent. Therefore, the main approach for generating a THz frequency comb has been to down-convert a near-infrared frequency comb by nonlinear frequency conversion using photomixing in photoconductive antennas[14] or optical rectification in crystals. Offset-free combs, of an average power usually limited to a few microwatts, are produced with the same line spacing as the near-infrared combs. The use of fibre lasers as near-infrared pumps and the design of novel photoconductive emitters with plasmonic enhancement[26] point to THz sources of smaller footprint and higher average power.

THz quantum-cascade frequency-comb lasers[27] hold promise for versatile and compact high-power electrically-pumped semiconductor sources that do not produce pulses, but still generate an array of a few hundreds of phase-coherent lines by four-wave mixing. With a careful design of the dispersion compensation, frequency-comb operation over a bandwidth close to an octave, around a central frequency of 3 THz (100µm), features [28] a power of 10 mW and a comb-line spacing of about 13 GHz.

Conversely, at large-scale facilities, synchrotron radiation in the so-called coherent mode shows temporal coherence, from one electron bunch to the next and over a revolution period in the storage ring [29]. This intriguing phenomenon remains to be exploited for spectroscopy.

2.4 Mid-infrared region.
Similarly, the mid-infrared region (2-20 µm, 15-150 THz) is technologically challenging. Ref. [30] reviews the field until 2012. Since then, direct mid-infrared generation of ultra-short pulses has progressed with e.g. a promising mode-locked $Er^{3+}$-doped fluoride glass fibre laser of 200-fs pulse duration[31] at 2.8µm (110 THz). As erbium, thulium and holmium present several gain bands in the mid-infrared region, further breakthroughs may be expected. Remarkable advances have also been achieved with new materials [32] and platforms for nonlinear frequency conversion and with quantum cascade lasers. Access to long wavelengths up to 12 µm (frequencies down to 25 THz) has become possible with nonlinear crystals such as orientation-patterned gallium phosphide [33] or $LiGaS_2$ [34]. Microresonators [35, 36] and waveguides [37-39] generate or broaden mid-infrared spectra, with on-chip silicon or silicon-nitride platforms, as reviewed in Ref. [40]. Quantum-cascade [41] and interband-cascade [42] laser frequency combs of GHz line spacing cover several spectral bands between 3 and 9 µm (33-100 THz) at an average power up to the Watt level, with about 100-200 comb lines. As quantum cascade devices also operate as detectors, they show an intriguing potential for fully integrated broadband spectrometers.

2.5 Visible and ultraviolet regions.
The visible and near-ultraviolet regions, down to wavelengths of about 200 nm (1,500 THz), are conveniently accessible by frequency-comb techniques using spectral broadening of near-infrared lasers in nonlinear waveguides and/or

harmonic- or sum-frequency conversion in nonlinear crystals. New crystals, such as $KBe_2BO_3F_2$ or $Li_4Sr(BO_3)_2$ [43], may extend the range down to 160 nm (up to 1,870 THz). Reaching shorter wavelengths involves high-harmonic generation in a rare gas, a very non-efficient process. For generating frequency combs of high repetition frequency (>20 MHz) suited for direct frequency comb spectroscopy in the extreme ultraviolet, the approach has been to inject the equidistant modes of an infrared, or visible, frequency comb into a resonant passive cavity containing the focus for the gas target [44, 45]. After each pass through the focus, the non-converted portion of the light pulse is coherently overlapped with the successive pulse from the laser. In this way, the intensity enhancement needed for high-harmonic generation can be reached. The approach is complex, as it also requires suitable out-coupling of the ultraviolet light and optimization of phase-matching effects that control the build-up of the harmonic signal over the interaction length. Recently, a record average power of 0.7 mW for a harmonic at 63 nm (4,760 THz) has been reported [46] at a repetition frequency of 77 MHz, using a swept cavity.

## 3. Spectrometric techniques for frequency comb spectroscopy

In most cases, the frequency-comb generator is a broadband light source that simultaneously excites several (many) transitions of the sample. Therefore, a spectrometer is needed, except for limited comb spans and/or very simple spectra. If the spectrometer has sufficient resolution, the individual comb lines may be resolved, enabling self-calibration of the frequency scale. Then the resolution is determined by the comb repetition frequency $f_{rep}$, although the spectral elements may be defined with a significantly higher precision. Once the resolution of the spectrometer is equal to – or better than – the comb line spacing, the instrumental line-shape, that convolves the atomic or molecular transitions, becomes determined by the width of the individual comb lines rather than by the spectrometer response. With a comb of narrow lines, the contribution of the instrumental line-shape becomes negligible when the atomic or molecular resonances have a width similar to or broader than the line spacing $f_{rep}$ of the comb. Before the advent of frequency comb spectroscopy, the instrumental line-shape in multiplex or multichannel spectroscopy used to be, at best, of similar width as the transitions. Furthermore, assuming that the sample does not change with time, the resolution may be enhanced, fundamentally down to the intrinsic comb line-width, by interlaying several spectra recorded e.g. with a tuned comb offset frequency $f_0$. We survey below the specific features of the spectrometric approaches that have been developed and exploited.

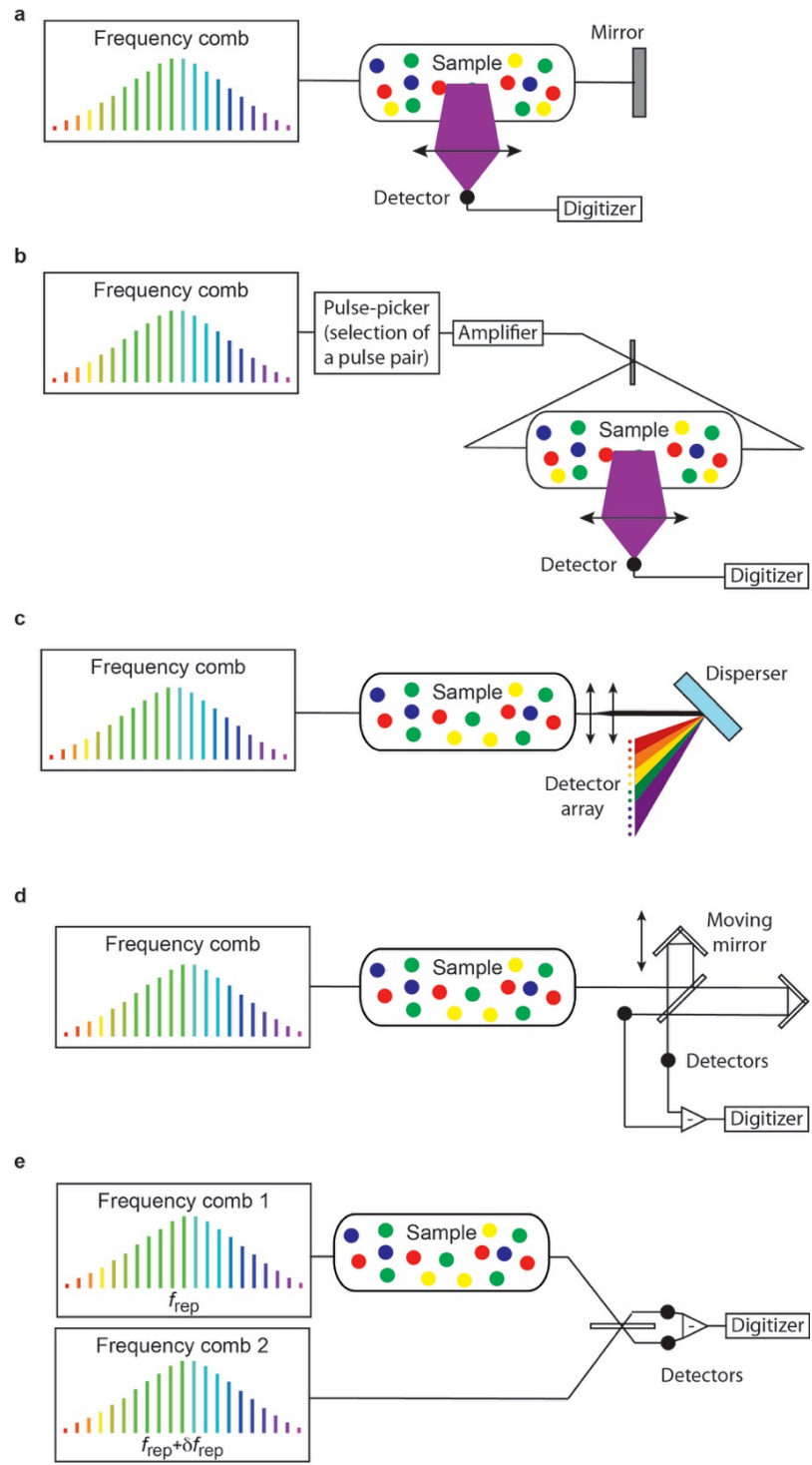

**Figure 3. Spectrometric techniques for frequency comb spectroscopy.**
**a.** Direct frequency comb spectroscopy with the example of two-photon Doppler-free excitation in a standing wave and detection of fluorescence of the sample.
**b.** Ramsey-comb spectroscopy also with the example of two-photon Doppler-free excitation in a standing wave and detection of fluorescence of the sample.
**c.** Frequency-comb spectrometry with a disperser for absorption measurements. Here a simple grating and a detector array are represented.
**d.** Frequency-comb Fourier transform spectroscopy with a scanning Michelson interferometer and an absorbing sample.
**e.** Dual-comb spectroscopy with one comb interrogating the sample and the other acting as a local oscillator. The absorption and the dispersion of the sample are measured.

3.1 Direct frequency comb spectroscopy
Direct frequency comb spectroscopy [8] (Fig. 3a) is the simplest approach to linear or nonlinear frequency comb spectroscopy. For linear spectroscopy [20, 47], a single comb line is resonant with a transition and all the other lines should ideally be detuned from resonances. For two-photon excitation [8, 20, 48], many pairs of comb lines of the same sum frequency contribute to the excitation (Fig.4a), which can be as efficient as with a continuous-wave laser of the same average power. The excitation can even be Doppler-free if the atoms are excited by two counter-propagating pulses forming a standing wave. The different schemes of two-photon excitation include stimulated Raman effects [49]. The response of the sample, e.g. its transmission, its fluorescence or its ionization rate, is recorded using a single detector. Sweeping e.g. the comb carrier-envelope frequency $f_0$ scans the spectrum, which is measured with a free-spectral range (Fig.5a) equal to the comb line spacing $f_{rep}$, ideally large. The approach is therefore only suitable for simple spectra comprising a few narrow transitions within the range of excitation and it has consequently been limited to gas-phase atomic systems. The technique is powerful, though: it is simple to implement compared to techniques requiring a spectrometer; absolute frequency calibration is obtained through knowledge of the repetition frequency $f_{rep}$ and of the carrier-envelope offset frequency $f_0$; as frequency combs often involve intense ultra-short pulses, nonlinear frequency conversion may be efficient and allows interrogating transitions in spectral ranges that are difficult or impossible to access with continuous-wave lasers. Pulse shaping allows to reduce the residual Doppler effect and to excite different transitions at distinct spatial locations [50]. For nonlinear excitation in the vacuum ultraviolet, reaching sufficient power with high-harmonic comb generators of large line spacing is a major challenge [46].

3.2 Ramsey-comb spectroscopy
Ramsey-comb spectroscopy [51] is a related time-domain technique (Fig.3b) that measures the interference between the excitations of an atomic or molecular sample by two time-delayed intense pulses derived from a frequency comb. The fringes, which are sensitive to the phase of the second pulse relative to the atomic coherence, are sampled at a set of different delay times, which are integer multiples of the pulse spacing of the laser oscillator plus some chosen fractional increments. The frequency of the excited transitions is deduced from fits of the portions of the phase signals. Similarly to direct frequency comb spectroscopy, the free spectral is limited to the comb repetition frequency $f_{rep}$, so that the technique is mostly suitable for metrology of simple spectra with few transitions. As already demonstrated in the deep ultraviolet with two-photon transitions of $H_2$ (Fig.5b) around 202 nm (1,485 THz) [52], Ramsey-comb spectroscopy holds particular promise, because the pairs of phase-coherent infrared pulses amplified to the millijoule level allow efficient frequency conversion.

3.3 Spectroscopy using a dispersive spectrometer
Dispersive spectrographs (Fig.3c) provide simple and robust tools for multichannel approaches to broadband spectroscopy with frequency combs. Gratings [15] and crossed dispersers, utilizing e.g. virtual imaging phase array (VIPA) étalons [16], have been successfully implemented with scanning single

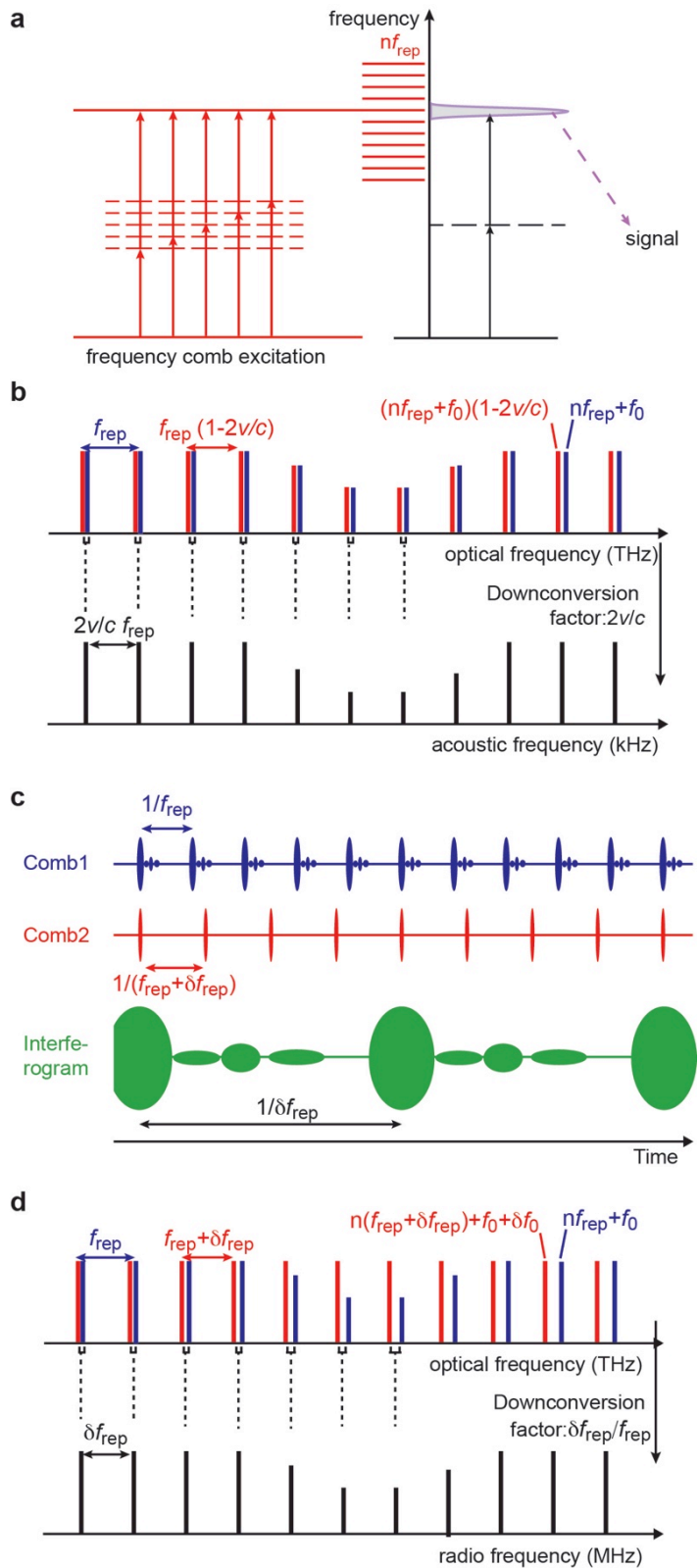

**Figure 4. Physical principle of some of the described spectrometric techniques**

**a.** In direct frequency comb spectroscopy with two-photon excitation, many pairs of comb lines may contribute to the excitation of the transition. However the spectrum is only measured modulo the comb repetition frequency. Fluorescence during decays towards lower energy levels may be detected.
**b.** In the moving arm of a scanning Michelson interferometer, the frequency of all the comb lines is Doppler-shifted. The beat notes between pairs of shifted and unshifted comb lines at the detector produce an acoustic comb.
**c.** Interferometric sampling in the time-domain stretches free-induction decay. With a dual-comb system, the interferogram recurs automatically at a period $1/\delta f_{rep}$ which is the inverse of the difference in repetititon frequencies of the two combs.
**d.** Frequency-domain picture of c for dual-comb interferometry. The beat notes between pairs of comb lines, one from each comb, generates a radio-frequency comb. The physical principle is the same as that of b., except that the down-conversion factor no longer depends on the speed of a moving part. Furthermore, dual-comb systems render the implementation of a dispersive interferometer easier.

detectors and with cameras. With a crossed disperser, resolutions as high as 600 MHz have been reported [53]. Most of the time, this is insufficient to resolve individual comb lines and many reports do not rely on the calibration by the frequency comb. Low-resolution spectrographs are however sufficient to resolve the individual lines of chip-based frequency comb generators such as microresonators. As advanced control and tuning of the line positions per steps across one free spectral range is feasible [54], rapid and compact instruments for gas-phase spectroscopy can even be envisioned. Alternatively, Fabry-Pérot filter cavities can increase the comb line spacing to exceed the spectrograph resolution. Vernier techniques [17], where the engineered mismatch between the free spectral range of a scanning Fabry-Pérot cavity and the frequency-comb line spacing is chosen as a ratio m/(m-1) with m integer, considerably relax the constraints on the spectrograph resolution. The Fabry-Pérot resonators may present a high finesse of several thousands: therefore they may simultaneously serve as enhancement cavities for weakly absorbing samples. Thanks to the availability of mid-infrared cameras with a short integration time, cavity-enhanced frequency-comb spectroscopy with a VIPA has proven [55] a powerful tool for monitoring, with a time resolution of 10 μs, the kinetics of the gas-phase reaction between carbon monoxide and the hydroxyl radical and for observing the intermediate hydrocarboxyl radical.

3.4 Michelson-based Fourier transform spectroscopy
Fourier transform spectroscopy with a scanning Michelson interferometer has been one of the most successful spectrometric techniques over the past fifty years. Usually associated with an incoherent broadband light source, the spectrometer [56] measures on a single photo-detector the interference between the two optically delayed signal from the two arms as a function of the optical path difference. The spectrum is the Fourier transform of the time-domain interference waveform, the interferogram.

Fourier transform spectrometry makes the best use of the available time and photons. It records spectra over extended spectral spans in any spectral regions and the spectral data, all simultaneously recorded, exhibit quality and consistency. Its instrumental line-shape is well understood and modelled. Its limitations are related to the presence of moving parts: the resolution is inversely proportional to the excursion of the moving arm and high-resolution instruments, commercially available with resolutions as high as 30 MHz, are slow and bulky.

In a Michelson interferometer, the frequency $f$ of the light traveling in the moving arm acquires a small Doppler-shift, equal to $-2 f v/c$, where $v$ is the velocity of the mirror, $c$ is the speed of light. At the output of the interferometer, the two electric fields coming from the fixed and moving arms beat on a photo-detector. The detector signal comprises this interference pattern, resulting from the down-conversion of the optical frequencies mostly to the audio-range. When a frequency comb is used as a light source in front of the interferometer (Fig.3d), the frequency of each comb line, reflected in the moving arm of the interferometer, is shifted (Fig.4b) by a factor $-2(n f_{rep} + f_0) v/c$, which gives the frequencies of the acoustic comb generated at the detector. The use of the frequency-comb synthesizer [19] has shown to add significant improvements to Fourier transform spectroscopy. A coherent light source such as a laser frequency comb has significantly higher brightness, leading to increased signal-to-noise ratio or decreased measurement times. Resolving the comb lines brings instrumental line-shapes of negligible contribution and direct calibration of the frequency scale (assuming that the comb is self-referenced), whereas accurate line position measurements previously relied on the presence of reference lines and careful assessment of the systematic effects. The frequency comb enables the implementation of detection techniques that straightforwardly retrieve the dispersion spectrum, leading to the measurement of real and imaginary part of the refractive index. The development of the technique has been fast, as it has profited from existing advanced techniques of control of scanning interferometers. Its use has been so far restricted to gas-phase spectroscopy in the infrared region. The sensitivity has been enhanced by synchronous detection [19], multi-pass cells [19] or high finesse cavities [57]. Cavity-enhanced frequency-comb Fourier transform spectroscopy has shown remarkable results for disentangling complex spectra (Fig.5c) of heavy molecules [58, 59].

### 3.5 Dual-comb spectroscopy

Dual-comb spectroscopy (Fig.3e) is a comb-enabled approach to Fourier transform interferometry without moving parts [60]. This instrumental scheme of frequency comb spectroscopy is currently that which attracts the highest interest. In most of the implementations, a frequency comb, of repetition frequency $f_{rep}$, interrogates the sample and beats on a fast photodiode with a second comb, of slightly different repetition frequency $f_{rep}+\delta f_{rep}$, which acts as a local oscillator. The interference signal is recorded as a function of time and it is Fourier transformed to reveal the spectrum. In the time domain (Fig.4c), the pulses from one comb walk through the pulses of the second comb with a time delay that automatically increases of an amount $\delta f_{rep}/f_{rep}^2$ from pulse pair to pulse pair. After interacting with the sample, the repeating waveforms of the electric field of the first comb are optically sampled by the local-oscillator comb and provide an interferogram stretched in time by a factor $f_{rep}/\delta f_{rep}$. In this way, optical delays between 0 and $1/f_{rep}$ are periodically scanned. In the frequency domain (Fig.4d), the beating signal of the two optical frequency combs of slightly different line spacing produces a comb in the radio-frequency domain that can directly be measured by digital electronics. The physical principle of the measurement is the same as that of the scanning Michelson-based Fourier transform spectrometer, with the practical difference that the down-conversion factor is not set by the speed of a moving part. It is therefore freely selectable within the Nyquist limit. The dual-comb spectrum may thus be mapped into the radio-frequency region, where the $1/f$ noise is reduced. A Michelson interferometer leaves to the user the choice of the resolution lower than the instrument capabilities: it allows any path difference excursions shorter than its maximum path difference, whereas the dual-comb spectrometer, by construction, automatically scans the optical delays up to $1/f_{rep}$. Even if numerical treatments, called apodization, make it possible to reduce the resolution, the time-domain scan always reaches a spectral resolution equal to the comb line spacing $f_{rep}$. If the desired resolution is significantly lower than $f_{rep}$, this results in wasted experimental time. Therefore, for optimized measurement times, dual-comb systems with a comb line spacing of the same order of magnitude as the desired resolution are advantageous in most cases. Conversely,

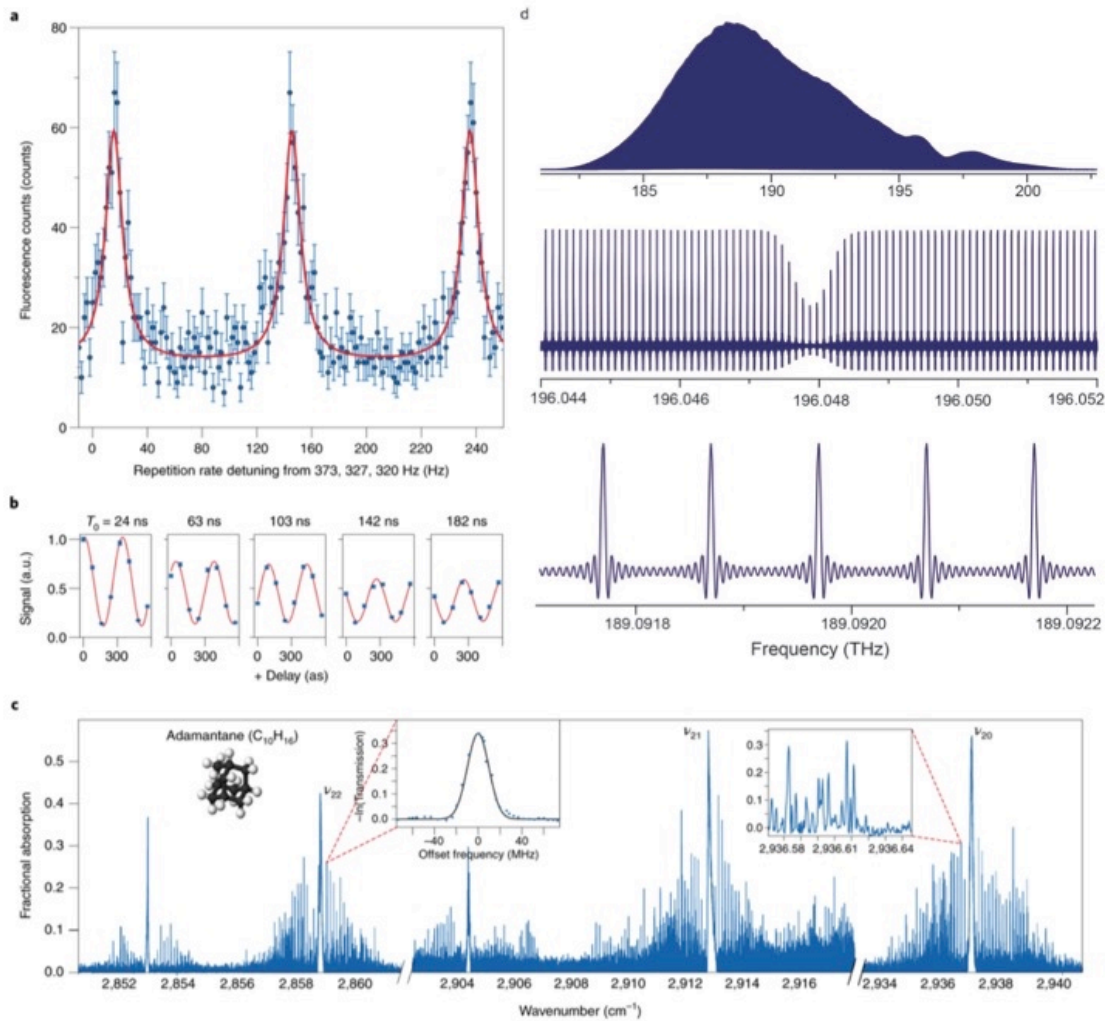

**Figure 5. Illustration of experimental results from the different approaches to frequency comb spectroscopy**

a. Direct frequency-comb one-photon spectrum [61] of the D2 line of a single $^{25}$Mg$^+$ ion around 280 nm (1,070 THz) observed through fluorescence. The spectral line recurs, as the free-spectral range is about 373 MHz.
b. Ramsey-comb interference fringes [52] of the $EF^1\Sigma^+_g$-$X^1\Sigma^+_g$ (0,0) $Q$1 rovibronic Doppler-free transition of H$_2$ excited by two photons at 202 nm (1,485 THz).
c. Cavity-enhanced frequency-comb Fourier transform spectrum [58] of buffer-gas cooled adamantine (C$_{10}$H$_{16}$) in the 3-µm region, at a translational temperature of 17 K and a spectral resolution of 10 MHz.
d. Near-infrared experimental dual-comb spectrum [62] showing 200,000 resolved comb lines with the expected cardinal-sine instrumental line-shape. The mutual coherence time of the interferometer is close to 2,000 seconds.

the resolution of spectra with resolved comb lines can always be improved by interleaving spectra [63].

A fundamental difference between dual-comb spectrometers and traditional ones is that the dual-comb approach is freed from geometry. With a dispersive or interferential instrument, the theoretical resolving power $R$ may be expressed as $R= \Delta/\lambda$, the ratio of the maximum path difference $\Delta$ to the wavelength $\lambda$. With a dual-comb interferometer, it becomes $R=T/\tau$ where $T$ is the measurement time and $\tau$ the period of the light vibration. Dual-comb spectroscopy is therefore the only technique that can, for any spans and any spacing, potentially reach a resolution equal to the comb line spacing. Such a specific feature may open up novel opportunities in precision spectroscopy and metrology, although it has not been exploited yet because of the technical challenges summarized below.

In addition to the challenges associated with the availability of the laser sources in the spectral regions of interest, a specific difficulty has slowed the applications of dual-comb spectroscopy to spectroscopic studies. The field has long been dominated by preliminary proofs-of-principle with a variety of original laser systems. The efficient measurement of interference requires time-domain coherence between the two interfering electric fields. Small relative timing- and phase-fluctuations between the two combs appear stretched by the same factor as the optically sampled waveform. In the frequency domain, the width of the beat notes between pairs of comb lines, one from each comb, needs to be narrower than the spacing of the radio-frequency comb lines, $\delta f_{\text{rep}}$, to preserve resolution. Furthermore, their intrinsic width should be narrower than the inverse of the measurement time, for best signal-to-noise ratio. The constraint is the same as that which require the interferometric control of the path difference in a Michelson-based Fourier transform spectrometer, technically mastered for decades. In dual-comb interferometry, the powerful yet simple techniques for referencing the comb to a radio-frequency clock, which are widely adopted in frequency metrology, do not provide the required short-term relative stability. Significant instrumental research has been undertaken to maintain, or reconstruct, the coherence in a dual-comb interferometer.

One approach has been to experimentally achieve such a mutual coherence by sophisticated servo controls. Locking two lines of each comb to a pair of narrow line-width continuous-wave lasers has yielded a coherence time inversely proportional to the line-width of the continuous-wave lasers. Using this principle, continuous time-domain measurements on the order of 1 second can be performed [64, 65]. Numerical techniques known as phase correction, derived from Michelson-based Fourier transform spectroscopy, can then be applied to efficiently average many recordings of 1 second and improve the signal-to-noise ratio. More recently, by feed-forward relative stabilization of the carrier-envelope offset frequency, the experimental coherence time reaches 2,000 s, without any indications that a limit is reached, in the near-infrared [62] (Fig.5d) and in the mid-infrared [66] regions. This suggests that, as with a Michelson interferometer, the experimental phase control of the dual-comb interferometer can be arbitrarily long, opening up novel opportunities for broadband frequency metrology.

Another approach has been to track the relative fluctuations between the two combs and to correct for these, either in real-time, by analog [67] or digital [68] processing, or *a posteriori* [69]. Such schemes have been implemented even with free-running or loosely locked lasers [67, 68]. They also compensate for the residual fluctuations of stabilized systems [70].

A third trend, which is currently stimulating many creative experiments, is to design dual-comb systems with built-in passive mutual coherence. Two trains of asynchronous pulses may be generated in dual-wavelength unidirectional mode-locked lasers [71], in bidirectional mode-locked lasers [72] and micro-resonators [73] e.g. by taking advantage of an asymmetry to nonlinear processes. A birefringent plate in a laser cavity has been harnessed [74] to produce overlapping cross-polarized pulse trains of a difference in repetition frequencies that depends on the optical thickness of the plate. The two combs of electro-optic-modulator based systems [24, 25, 75] can share a number of components. With some of these set-ups [75], the mutual-coherence time exceeds 1 second. Such designs without any stabilization electronics greatly facilitate dual-comb spectroscopy and hold promise for transportable or even portable compact and easy-to-use interferometers.

Most of the time, dual-comb interferometers are designed for linear absorption spectroscopy. When the sample only interacts with one comb, the phase spectrum is simultaneously obtained, whereas it was technically challenging to investigate with dispersive Michelson interferometers. Direct access to both the real and imaginary parts of the refractive indices is given. With the objective of linear absorption measurements, numerous demonstrations have been accomplished with a variety of laser sources, mostly across the infrared region (Fig.6); to cite just a few: near-infrared erbium-doped [18, 62, 64, 65, 67, 68, 71, 72] and ytterbium-doped [76] fibre combs, electro-optic modulators [24, 25], near-infrared [77, 78] and mid-infrared [79] microresonators, frequency-doubled fibre lasers [80, 81], mid-infrared [82] and THz [28, 83] quantum cascade or interband [42] cascade lasers, mid-infrared $Cr^{2+}$:ZnSe lasers [84], mid-infrared optical parametric oscillators [85-87], mid-infrared difference-frequency systems [12, 70, 75, 88], THz photoconductive antennas [14, 89, 90].

As dual-comb interferometers often harness ultra-short-pulse lasers, they enable novel nonlinear broadband spectroscopy. New schemes of coherent Raman spectroscopy (CARS) [91], stimulated Raman spectroscopy [92] and two-photon excitation [81] have been first demonstrated, followed by others such as pump-probe spectroscopy [93]. Sometimes, e.g. in dual-comb two-photon excitation[81] with background-free fluorescence detection, the interferometric modulation is measured indirectly, through its transfer to e.g. the modulation of the intensity of the fluorescence of the sample, providing an insightful illustration of Fourier encoding. The first proof-of-concept [94] of broadband Doppler-free two-photon spectroscopy showcases experimental spectra spanning 10 THz, which exhibit atomic line profiles of a width of 6 MHz.

A strength of dual-comb spectroscopy (and other techniques of Fourier transform spectroscopy), that has been recently explored, is its ability to efficiently combine with other sampling techniques or instrumentations. The sensitivity to weak absorption is enhanced with multipass cells and enhancement cavities [76, 95]; the signal-to-noise ratio is increased by electro-optic sampling [89]; spectro-imaging [91] and microscopy [96] provide spectral maps of spatially inhomogeneous samples. Some selected examples are highlighted in section 4. The attractive advantage of multiplex measurements over extended spectral spans, which provides overall consistency and short measurement times, is added to those of the sampling techniques. Specific features of dual-comb systems include the absence of moving parts, the use of laser beams rather than incoherent light, the feasibility of short measurement times, the absolute frequency calibration and the achievable negligible contribution of the instrumental line-shape.

### 3.6 Other approaches
Alternate methods, involving e.g. speckles in multimode fibres [97], sweeping the comb repetition frequency to generate a time delay between the two arms of a static interferometer [98], cavity filtering and scanning [99] or heterodyning a comb with a continuous-wave laser [100] have been explored. Though they have not been widely adopted yet, they may be particularly useful in some circumstances.

## 4. Selected applications and prospects.

Most techniques of frequency comb spectroscopy are recent, thus their applications are only in their infancy. We briefly review here a selection of those and discuss the envisioned trends and prospects.

### 4.1 Precision spectroscopy: towards the extreme ultraviolet and broadband detection?
Precision spectroscopy of atomic transitions has been the most investigated application of frequency comb spectroscopy. Such measurements enable stringent tests of fundamental theories, accurate determinations of physical constants, and searches for new physics. With direct frequency comb spectroscopy or Ramsey-comb spectroscopy, the absolute frequency of narrow transitions in atomic systems is determined, such as in argon [47], cesium [51], krypton, hydrogen [48], magnesium, single magnesium ion[61], neon [47], single calcium ion [49], rubidium [51] in gas cells, ion traps, laser-cooled systems or atomic beams. Only one molecule, $H_2$, has been considered so far [52]. The fractional uncertainty of the frequency measurements is typically on the order of $10^{-10}$, but it can reach [48, 51] parts in $10^{12}$. The extension of the techniques of precision spectroscopy with frequency-comb excitation to the extreme ultraviolet, where continuous-wave lasers are not available, portends fascinating opportunities for fundamental physics, including better tests of quantum electrodynamics, of molecular quantum theory, or future nuclear clocks.
Moreover, the prospect of Doppler-free spectroscopy [94] or of cold-molecule spectroscopy [58] over broad spectral bandwidths is opening new perspectives to precision spectroscopy: detailed analysis e.g. of simple molecules with few electrons over an extended range may deliver new information on molecular structure and potential-energy surface and may help to validate or improve *ab initio* quantum-chemical computations.

### 4.2 Laboratory molecular spectroscopy over broad spectral bandwidths
Because of the instrumental challenges of frequency comb spectrometry, spectroscopic studies have remained rather scarce, but the few published ones are likely to stimulate further contributions from the growing community. Most of the spectroscopy work has been performed in the near-infrared region, where the technology is more mature. Mainly line positions and shifts [101] have been determined. The unique feature of frequency-comb spectrometers with resolved narrow comb lines, the negligible contribution of the instrumental line-shape, will permit the metrology of line parameters other than line positions: frequency comb spectroscopy combines for the first time a broad spectral span, which was the distinctive feature of spectrometers with incoherent light sources, and an narrow instrumental line-width, which used to be the specific character of some tunable lasers. New investigations for a better understanding and modelling of spectral line-shapes may be triggered. One of the first published studies [102] illustrates such benefits for the modelling of near-infrared spectra of water vapour at high temperature.

The combination of fast measurement times and (sometimes moderately) broad spectral bandwidth advances time-resolved spectroscopy, with time resolutions on the scale of several microseconds, in a variety of situations, from the investigation of chemical gas-phase reactions through mid-infrared spectroscopy [55] to the kinetics of spectral hole burning in a transition of atomic caesium [103]. Gas-phase transient absorption of electronic transitions of diatomic molecules in the visible range [104] and vibrational and electronic population relaxation of dye molecules in solution [105] associate frequency comb spectroscopy to the study of ultrafast phenomena.

The past few years have witnessed a diversification of the samples, which are no longer restricted to the gas phase. Samples in the liquid [91, 92, 96, 105-107] or solid [108] phases have been studied. With them comes the requirement of sources and spectrometric techniques suited to their broad spectral transitions and their extended spectra. The development of spectrometric techniques involving microresonator-based frequency combs of large line spacing, especially in the mid-infrared [79], is therefore very timely. Nonlinear spectroscopy with combs of high repetition frequency faces the specific difficulty [91] of a lower energy per pulse and requires novel strategies to be devised.

4.3 Coherent control and multi-dimensional spectroscopy
Laser frequency comb techniques can measure and control the phase of an optical electric field with respect to the corresponding intensity waveform [109]. This is opening up new horizons for the generation of arbitrary waveforms at optical frequencies and "line-by-line" pulse shaping. It might create novel opportunities for coherent control in chemical reactions. The intense ultra-short laser pulses of the frequency combs can be further harnessed to exploit complex pulse sequences and coherent transient phenomena including photon echoes, in analogy to multidimensional nuclear magnetic resonance spectroscopy. On a longer term, by rapidly exploring a multi-dimensional parameter space, nonlinear multi-dimensional multi-comb spectroscopy and imaging might reveal much additional information inaccessible by conventional linear and nonlinear spectroscopy. Two-dimensional spectroscopy of gas-phase alkali atoms has been explored with a dual-comb [110] system generating photon echoes, at spectral resolutions that would be difficult to reach with the mechanical delay lines commonly used in multidimensional spectroscopy. Theoretical proposals and insights may help guiding the vision [111].

4.4 Environmental sensing

Frequency comb spectroscopy presents some interesting characteristics for fieldable and even portable gas sensors. For instance, the laser beams enable long open-path propagation, filling the gap between point sensors and remote-sensing instruments, e.g. on board satellites, aircrafts or balloons. The spectra show high consistency, stability and repeatability for concentration measurements. The broad spectral bandwidth enables detection of multiple species, as well as more reliable inversion. The applications range from the monitoring of greenhouse gases to industrial process control or leak detections. Significant progress has already been achieved toward the objective of compact portable *in situ* frequency-comb spectrometers. A transportable dual-comb sensor, deployed in the field, shows [112] continuous monitoring and quantification of methane emission sources at a regional scale, with the prospect of efficient leak detection at oil and gas operation facilities. In a laboratory proof-of-principle of dual-comb spectroscopy of laser-induced plasmas [113], time-resolved broadband spectral analysis of laser-ablated solid materials is performed and lays the first bases for *in situ* laser-induced breakdown spectroscopy of solids, liquids and aerosols. The demonstrations have so far been accomplished in the near infrared region. Improvements to fibre-, semiconductor- and chip-based instrumentation will render the sensors more compact, rugged and easy-to-use, even in harsh environments. Continued progress to mid-infrared frequency-comb sensing technology will increase the number of detectable molecules, as well as the detection sensitivity.

4.5 Applications to chemistry, biology and medicine

Frequency combs will expand the capabilities of optical spectroscopy, spectro-microscopy and hyperspectral imaging for chemical or bio-medical analysis. Breath analysis by cavity-enhanced direct frequency-comb spectroscopy has also been envisioned [114]. An even more intriguing prospect is the potential of frequency combs for physical chemistry in condensed matter. Indeed, harnessing frequency combs for "low-resolution" spectroscopy may initially be seen as non-intuitive and thought provoking. However, converging insights and first proof-of-concepts provide a set of arguments. Chip-scale dual-comb spectrometers with mid-infrared combs of large line-spacing [79, 115] may bring new tools for time-resolved spectroscopy of samples in the condensed phase. First dual-comb spectrometers for spectro-imaging [91], confocal microscopy [96] and near-field microscopy [116] showcase a short measurement time per pixel and a high spectral resolution.

4.6 Towards a spectroscopy laboratory on a chip?

The large line spacing of frequency comb synthesizers such as microresonators [77-79] or quantum cascade lasers [82, 83] renders them particularly suited for real-time vibrational dual-comb spectroscopy, especially in the mid-infrared region. With continued technology progress, significantly broader spectral spans than currently reported may be achieved. In dual-comb spectroscopy, a comb line spacing of the same order of magnitude as the resolution optimizes the measurement time [91]. Therefore frequency combs of large repetition frequency are pivotal to dual-comb spectroscopy in condensed matter. For instance, combs with a line spacing of about 20 GHz and a span of 20 THz would in principle allow the recording of dual-comb spectra of 20-GHz resolution at a refresh time of 100 ns (10 MHz). The usually poor dynamic range of multiplex techniques, combined with very-short detection times, may thus create significant technical difficulties, which might be overcome with clever pulse shaping or spectral filtering. Despite the daunting challenges related to the detection sensitivity, the prospect of a new tool for single-shot time-resolved multiplex spectroscopy is very appealing. In a long-term vision, the frequency-comb sources, the sample interrogation (which may include e.g. evanescent sensing as already demonstrated with a dual-comb system in the gas phase [117]) and the detection system would be entirely integrated on a single chip-scale device for analytical chemistry.

## 5. Conclusion

Frequency comb spectroscopy is a recent field of research that has blossomed in the past five years. Extensive developments in laser technology have brought to fruition comb sources with specifications that often fulfill the experimental requirements of the various schemes of frequency comb spectroscopy. The last few years have witnessed remarkable advances in the mid-infrared molecular fingerprint region, which appear ready to benefit laboratory (ro-)vibrational spectroscopy and various applications to sensing. The (extreme) ultra-violet region is still technically challenging and has been less investigated, despite its interest for electronic spectroscopy of atoms and molecules with applications that range from fundamental physics and chemistry to laboratory spectroscopy in support to astrophysics. The performance frontiers of the comb-based instruments have been advanced quickly and, in many schemes, instrumental artifacts are understood, controlled and minimized. Broadband multiplex or multichannel spectra of the linear complex response of a sample become measurable across the THz, infrared and visible region with a frequency scale directly referenced to an atomic clock and an instrumental line-shape of negligible contribution for transitions broadened by collisions or by the Doppler effect. Novel techniques of nonlinear spectroscopy and of multidimensional spectroscopy promise new insights into the structure of matter and its structural change and coupling in molecular dynamical processes. Frequency comb spectroscopy begins to investigate questions that had not been foreseen twenty years ago. The efforts are now shifting toward the physics, the chemistry, and maybe even the biology, that can be explored with the new tools. Recent reports indicate promising directions: with disentangled complex molecular spectra and multiplex Doppler-free spectroscopy over a broad spectral bandwidth, precision spectroscopy may broaden its scope and provide new tests of fundamental physics and chemistry. Spectroscopy of single events, kinetics and microscopy with rapid spectrometers with capabilities of time-resolution and spatial resolution may find applications in chemistry and biology. Nano-photonics takes the instrumentation to the chip scale. We hope that this short account of the rapidly evolving field of frequency comb spectroscopy will help stimulate further developments of creative instrumentation and techniques, and that it will inspire future applications that take advantage of some of the emerging unique capabilities.

**Acknowledgments.** Support by the Carl-Friedrich-von-Siemens Foundation is gratefully acknowledged.

**Correspondence.** Correspondence should be addressed to Nathalie Picqué (nathalie.picque@mpq.mpg.de)